\documentclass[twocolumn]{aastex63}
\usepackage{xcolor}

\accepted{9 Dec 2021}
\submitjournal{ApJL}

\shorttitle{}
\shortauthors{Lintott, Bannister \& Mackereth}

\newcommand{\okina}{\textquoteleft}

\newcommand{\Ou}{{\okina}Oumuamua}

\graphicspath{{./}{figures/}}

\begin{document}

\title{Predicting the water content of interstellar objects from galactic star formation histories}





\author[0000-0001-5578-359X]{Chris Lintott}
\affiliation{Department of Physics, University of Oxford, Denys Wilkinson Building, Keble Road, Oxford, OX1 3RH, UK}
\author[0000-0003-3257-4490]{Michele T. Bannister}
\affiliation{School of Physical and Chemical Sciences --- Te Kura Mat\={u}, University of Canterbury,
Private Bag 4800, Christchurch 8140,
New Zealand}
\author[0000-0001-8108-0935]{J. Ted Mackereth}
\altaffiliation{Banting Fellow}
\affiliation{Canadian Institute for Theoretical Astrophysics, University of Toronto, 60 St. George Street, Toronto, ON, M5S 3H8, Canada}
\affiliation{Dunlap Institute for Astronomy and Astrophysics, University of Toronto, 50 St. George Street, Toronto, ON M5S 3H4, Canada}
\affiliation{David A. Dunlap Department for Astronomy and Astrophysics, University of Toronto, 50 St. George Street, Toronto, ON M5S 3H4, Canada}

\begin{abstract}

Planetesimals inevitably bear the signatures of their natal environment, preserving in their composition a record of the metallicity of their system's original gas and dust, albeit one altered by the formation processes. When planetesimals are dispersed from their system of origin, this record is carried with them. As each star is likely to contribute at least $10^{12}$ interstellar objects, the Galaxy's drifting population of interstellar objects (ISOs) provides a overview of the properties of its stellar population through time. Using the EAGLE cosmological simulations and models of protoplanetary formation, our modelling predicts an ISO population with a bimodal distribution in their water mass fraction: objects formed in low-metallicity, typically older, systems have a higher water fraction than their counterparts formed in high-metallicity protoplanetary disks, and these water-rich objects comprise the majority of the population. Both detected ISOs seem to belong to the lower water fraction population; these results suggest they come from recently formed systems. We show that the population of ISOs in galaxies with different star formation histories will have different proportions of objects with high and low water fractions. This work suggests that it is possible that the upcoming Vera C. Rubin Observatory Legacy Survey of Space and Time will detect a large enough population of ISOs to place useful constraints on models of protoplanetary disks, as well as galactic structure and evolution.

\end{abstract}

\keywords{ Interstellar objects (52), Galaxy Evolution (594) }

\section{Introduction} \label{sec:intro}

The discovery of 1I/\Ou\ \citep{Meech2017} and 2I/Borisov\footnote{\url{https://minorplanetcenter.net/mpec/K19/K19RA6.html} and \url{https://minorplanetcenter.net/mpec/K19/K19S72.html}}, the first two interstellar objects detected while passing through our Solar System, has stimulated interest in the properties of the population from which they are drawn. Such small bodies are likely to be produced by a wide variety of dynamical and physical processes, in common with the development of planetesimals \citep{ISSITeam:2019}. There is strong evidence that the early history of planetary systems, including our own, includes a period where $\sim 90$\% of a system's planetesimals are ejected \citep{Fernandez:1984, Izidoro:2015, Raymond:2020}. Most of the remaining planetesimals are also expected to become unbound from their parent star in time, through stripping by the Galactic tide, stellar encounters, and in the post-main-sequence stages of stellar evolution \citep{Kaib:2009, Veras:2011, Veras:2016, Pfalzner:2021flyby}.

In this work, we view these dispersed planetesimals as forming a population of interstellar objects (ISOs) with the potential to encounter our Solar System. 
Each stellar system contributes to a progressively accumulated background population of ISOs that builds up through Galactic history \citep{MoroMartin:2009, MoroMartin:2018, Pfalzner:2019}, and which is then sampled as the Solar System moves through the Milky Way. While more exotic processes may also contribute to the ISO population (see \citet{Levine:2021} for an overview of current hypotheses), we address only planetesimal contributions in this initial study.

The properties of the interstellar objects that pass through our Solar System are primarily determined by the material and conditions in the protoplanetary disks in which they formed. The dependence of the water abundance in protoplanetary bodies on the composition of their parent star has long been acknowledged, with the relative abundance of carbon and oxygen of particular importance \citep{DelgadoMena:2010,Johnson:2012}. More recent work by \citet{Cabral:2019} developed a model of the expected composition of planetary building blocks on stellar metallicity; similar work by \citet{Santos:2017} considered the dependence of water fraction of assembled, mature planets on initial metallicity. The model developed by \citet{Bitsch:2020} (hereafter BB20), discussed in detail below, also indicates that the properties of planetesimals are strongly dependent on disk metallicity. In particular, the BB20 model suggests that the water content of planetesimals that form beyond the ice line strongly depends on the availability or otherwise of oxygen.

Therefore, if we want to make predictions about observable quantities, such as the water content of ISOs such as \Ou\ and 2I/Borisov, we need to consider the properties and metallicity of the protoplanetary disk population in the Milky Way. This paper is a first attempt to combine existing models for this purpose. 

Luckily, enormous progress has been made in recent years in producing models of the structural and chemical evolution of the galaxy (see, for example, reviews by \citet{Bland-Hawthorn:2016} and \citet{Helmi:2020}). Galactic models are increasingly well constrained by data, particularly from Gaia \citep{Gaia:2018} and ground-based spectral surveys such as GALAH \citep{Buder:2018} and APOGEE \citep{Majewski:2017}. Detailed simulations which put galactic structure in a cosmological context have been developed, allowing the chemical evolution of components in Milky Way-analogue galaxies to be studied in detail \citep[e.g.][]{Schaye:2015, Crain:2015, Wetzel:2016, Nelson:2019, Pillepich:2019}.

In this paper, we use the EAGLE suite of hydrodynamical cosmological simulations \citep{Schaye:2015, Crain:2015}, alongside the models of \citet{Bitsch:2020}, to explore the connection between galactic chemical evolution and the composition of the observed ISO population for the first time. This work demonstrates the interest and importance of this chemical evolutionary approach to predicting the properties of the interstellar object population, and conversely, to show how the properties of observed ISOs might constrain the Milky Way's star formation history. Existing constraints on such models come primarily from observations of stellar abundance. Though these observations are clearly linked to the metallicity of the protoplanetary disk, thermal and chemical reprocessing will affect how the relative abundances of species within a protoplanetary disk differ from its natal molecular cloud \citep{Fedele:2020}. It is therefore possible that studies of large numbers of interstellar objects may eventually provide insight into such processes. 

In this Letter, we briefly review our input models, before deriving a prediction for the distribution of water content of the predicted interstellar object population. With a larger local population expected to be explored by the first few years of the Vera C. Rubin Observatory's upcoming Legacy Survey of Space and Time (LSST) \citep{Ivezic:2019}, we hope that this paper is the beginning of an effort to place interstellar objects passing through our Solar System in the context of Milky Way models. 

\section{Input models}
\subsection{Planetesimal formation and composition}
\label{sec:BBmodel}

The detection of \Ou\ requires that a large population of ISOs must exist. 
For instance, \citet{Meech2017} consider this together with the non-detections in a volume of thoroughly searched survey space over 17 years to constrain the local density of similar ISOs as $10^{15}\ \mathrm{pc}^{-3}$. \citet{Do:2018} similarly find a mass density of ISOs that is high enough to imply that every star is contributing to the population. The discovery of 2I/Borisov is consistent with a size distribution for these objects that implies that ISOs are common in our solar neighbourhood \citep{Jewitt:2019b}. As discussed above, the origin of the bulk of this population will be from planetesimal formation, and the properties of such objects must depend on conditions in the protoplanetary disk, and in particular on the raw materials available. 
The connection between stellar metallicity and the presence and properties of planetary systems has been debated since the discovery of the first exoplanets; a connection between metallicity and the probability that the star hosts giant planets was first suspected when only a handful were known \citep{Gonzalez97} and was quickly formalised.  \citet{FischelValenti:2005} found that the formation probability for gas giant planets is related to the square of the metallicity. It may even be possible to predict which stars have planets based on their metallicity \citep{Hinkel:2019}. 

Recent work by \citet{Sousa:2019} found, furthermore, that properties of planets, specifically their mass, may itself depend on stellar metallicity. Models of planet formation need, therefore, to consider the composition of accreting material. While many models follow astronomical convention in using the stellar iron abundance ([Fe/H]) as a single metallicity parameter, scaling other atomic species as appropriate, we adopt the models of \citet{Bitsch:2020} which consider separately abundances for several important elements: iron, silicon, magnesium, oxygen and carbon. 

In this paper, as described below, we will assume that the composition of the interstellar object population is determined by that of the parent population of protoplanetary objects. We will use the models of \citet{Bitsch:2020}, and refer the interested reader to the detailed description of their work. In brief, we should note that their model derives from stellar abundances from the GALAH survey \citep{Buder:2018}, which allows the relationship between overall metallicity parameterized by [Fe/H] and the abundance of other elements to be measured. The authors then use a chemical model derived from \citet{Madhusudhan:2017}) and \citet{Bitsch:2018} to predict the abundance of molecules and resulting composition of planetary building blocks formed on either side of the water ice line\footnote{The mode of planetesimal formation followed depends on whether the disk is locally cold enough to allow water ice to form.}. For this initial work, we assume that the ISO population is exclusively drawn from planetesimals which are formed beyond the ice line. This assumption is consistent, for instance, with ISOs liberated from their systems of formation by intracluster stellar flybys \citep{Pfalzner:2021flyby}, though it will not fully account for the effect of planetary migration and scattering; however, similar trends in composition are seen by BB20 either side of the ice line. 

The outputs of the model are shown in Figure 10 in \citet{Bitsch:2020}. Though a variety of molecules show changes in mass fraction beyond the ice line, we concentrate here on the mass fraction in water, which shows the most significant decline in mass fraction as stellar metallicity increases. 


While at low metallicities water accounts for roughly half the mass in the planetesimal population, as the metallicity increases it becomes much less significant, with the formation of $\mathrm{CO_2}$, $\mathrm{CH_4}$ and $\mathrm{CO}$ through chemical pathways which compete with water formation favoured instead. The mass fraction of water thus declines with metallicity. 

For our study, we read predicted mass fractions for [Fe/H] of -0.4 to 0.4 in intervals of 0.1 from Figure 10 in BB20, interpolating between the plotted values using a third-degree polynomial fitted to the data. We will need to extend the model to higher metallicity for the most recently formed stars. In these regimes we extrapolate beyond the limits of the BB20 models and treat the water mass fractions there as upper/lower limits. In practice we need only assume that the water mass fraction stays low beyond [Fe/H]=0.4; this is justified as BB20 point out that increasing [C/O] with [Fe/H] in Milky Way stars indicates that water mass fractions should continue to decrease with metallicity, and the expected trend continues well outside the BB20 [Fe/H] range in data from the APOGEE DR16 results \citep{Majewski:2017, Ahumada:2020}.

\subsection{Chemical Evolution Models}
\label{sec:galmodel}
Understanding the distribution of chemical elements throughout galaxies is essential in trying to understand the formation history and subsequent evolution of such systems. Measurements of stellar metallicity from surveys such as APOGEE \citep{Majewski:2017} which include hundreds of thousands of stars provide a `fossil record' for our own galaxy. Such a record is necessarily affected by a strong selection function (APOGEE, for example, targets only red giants and obviously cannot observe massive stars which have already died), and so the data is interpreted with the help of cosmologically motivated, hydrodynamical simulations such as EAGLE \citep{Schaye:2015,Crain:2015}, which predict the star formation histories and evolution of many hundreds or thousands of systems, providing a historical record of their evolution which can be compared with APOGEE to identify Milky Way-like systems. As such simulations contain details of every epoch of star formation, rather than just those whose stars survive to the present day, they are invaluable in allowing us to derive from simulations a history of star formation for a galaxy like our own. In this paper, we will assume that ISOs are produced along with stars, inherit a composition appropriate to their metallicity (see previous section), and remain in the wandering population even when the stars that produced them have reached the end of their lives. We further assume that the population is mixed throughout the Milky Way, so that any star formed can contribute to the observed ISO population; more complex modelling of galactic dynamics is left for future work. 

We begin by using a wide selection of galaxies from the Ref-L100N1504 simulated volume to explore how our predictions depend on the star formation history of the simulated galaxy, selecting all 2039 galaxies from the Ref-L100N1504 volume which have $10^{10} < M_{\star} < 10^{11}\ \mathrm{M_{\odot}}$. 

We next follow \citet{Mackereth:2018} (hereafter M18), who investigate the chemical evolution of more than a hundred Milky Way-like galaxies drawn from EAGLE (specifically, from the Ref-L100N1504 simulation). Galaxies included in the study were required to have stellar mass in the interval $\mathrm{M_*}=(5-7)\times 10^{10}\mathrm{M_(\odot)}$ and to be disk dominated\footnote{The latter selection was carried out by requiring the fraction of kinetic energy carried by particles participating in ordered rotation to be greater than 0.4.}. 

Finally, we use data from a single simulated galaxy which, like the Milky Way, exhibits $\mathrm{[\alpha/Fe]}$ bimodality (at fixed $\mathrm{[Fe/H]}$) and a star formation history which allowed high- and low-$\mathrm{[\alpha/Fe]}$ populations to evolve separately. This combination of properties is rare --- only 6 of the 133 systems which meet the Milky Way like criteria used by M18 show similar features --- and makes the system a good analogue for our own galaxy. The complex history of gas accretion and star formation which takes place in the simulated galaxy, and the determination of the properties described above, are discussed in detail by M18 in their Section 4. 

Within the simulation, gas particles undergo star formation via a stochastic process, with a probability that depends on the star formation rate and the gas particle mass (see \citet{Schaye:2015} and \citet{Crain:2015} for details of the treatment of star formation and enrichment). When it occurs, star formation follows a pressure dependent Kennicutt-Schmidt law, and is assumed to produce a single stellar population with a Chabrier IMF \citep{Chabrier:2003}. We assume that a population of ISOs is created at the moment of star formation, though, in practice, there will be a delay between the formation of the star and the creation of planetesimals and their expulsion from the system. As the majority of such expulsions will happen early in a star's life \citep{Pfalzner:2019}, we do not attempt to model this separately. We assume that the number of ISOs contributed to the background population is proportional to the mass of the star. This assumption is justified as the mass of observed protoplanetary disks scales in a roughly linear fashion with the mass of the star \citep{Andrews:2013}. Large ISOs will be long-lived, and so once added to the galactic population they will persist \citep{Guilbert-Lepoutre:2016}. 

\section{Results}
\label{sec:results}

\subsection{Example systems}
\label{sec:examples}
We begin by using the mass and [Fe/H] data provided by the simulations for each star particle in two example galaxies and the models from BB20 described in section \ref{sec:BBmodel} to predict the water abundance of the ISOs in each system. We choose to focus on water as an abundant molecule, and because the models of BB20 also show a much greater variation in water abundance with metallicity than other molecules. The water fraction of any observed and suitably bright ISO may potentially be determined from their comae, meaning that our model could be tested as more interstellar objects are found. By comparing two galaxies, we hope to illustrate the sensitivity of the ISO population's predicted water-rich fraction to the details of their star formation histories. A large range in the predicted value would indicate that any model which predicts the observable properties of the ISO population would need to take into account the specific star formation history of the Milky Way. It would also suggest that observations of a significant number of ISOs might provide an independent observational test of the Milky Way star formation history.

\begin{figure}
    \centering
    \includegraphics[width=\columnwidth]{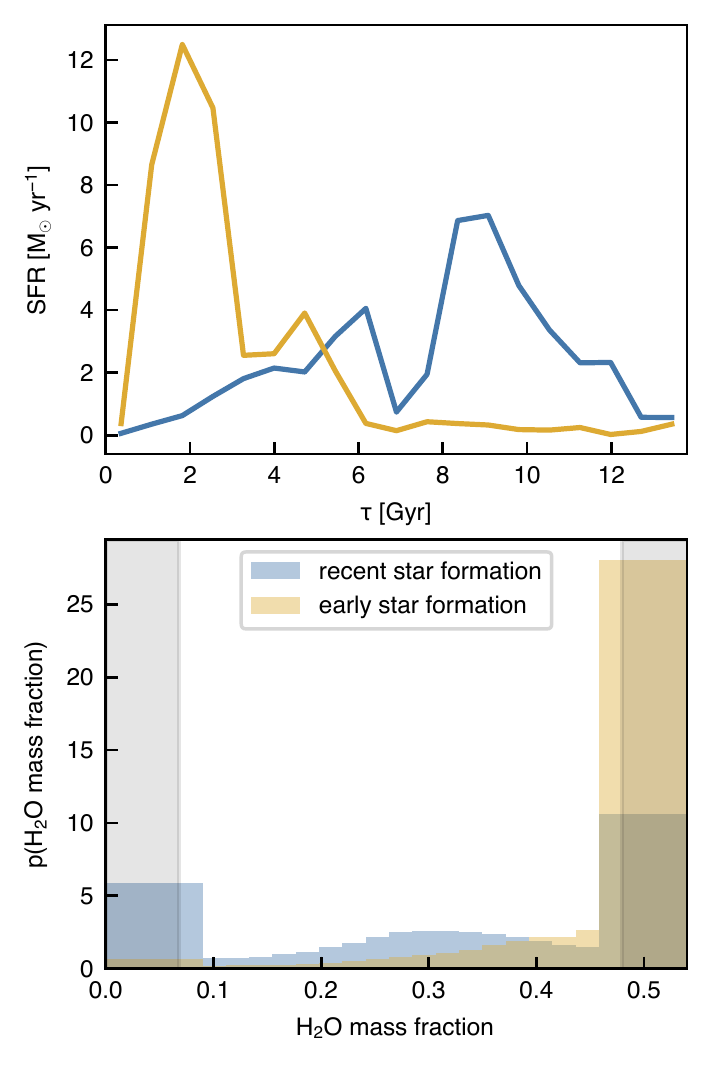}
    \caption{Top: Star formation histories of two example galaxies (at fixed stellar mass) selected from EAGLE to demonstrate the ISO population predicted in a galaxy with recent star formation (blue) vs. one with earlier star formation (yellow). Bottom: Predicted water mass fractions of ISO populations predicted for the galaxies shown above. The histograms are normalised such that $p(\mathrm{H_2O\ mass\ fraction})$ is the probability density for finding an ISO in a given mass fraction bin or equivalently the fraction of the predicted ISO population in each bin.}
    \label{fig:sfh}
\end{figure}

The two galaxies have $5 < M_{*} < 7\times10^{10}\ \mathrm{M_{\odot}}$. They differ in that one has star formation concentrated at early times, whereas the other has a more recent peak in star formation rate. Their star formation histories are shown in Figure \ref{fig:sfh}(a), and the fraction of the predicted ISO population with particular mass fraction in water in Figure \ref{fig:sfh}(b). The galaxy with early star formation has a ISO population dominated by water-rich ISOs formed in low-metallicity systems, whereas in the system where star formation peaks more recently we gain a low-water mass fraction population formed in higher metallicity systems.

For convenience later in the paper, we will quantify the differences in population by considering the the fraction of water-rich ($f_{\mathrm{H_2O}} > 0.4$) to water-poor ISOs. In the galaxy with only early star formation, this water-rich fraction is 0.74, whereas it is lower at 0.32 for the galaxy with recent star formation. The difference indicates that a galaxy's ISO population's properties --- at least in the case of the water mass fraction --- does indeed depend on its star formation history.

\subsection{Trends in the galaxy population}

To further illustrate the sensitivity of our results to Galactic history, we compare the predicted water-rich fraction of the observed ISO population in a wider sample of galaxies drawn from the EAGLE simulation. For each, we compute their ISO population and mean stellar metallicity, $\mathrm{[Fe/H]}$, as in Section \ref{sec:examples}. We then compute the fraction of water-rich ISOs (i.e. those with an $f_{\mathrm{H_2O}} > 0.4$) in each galaxy and compare this with the median age of its star particles.

\begin{figure}
    \centering
    \includegraphics[width=\columnwidth]{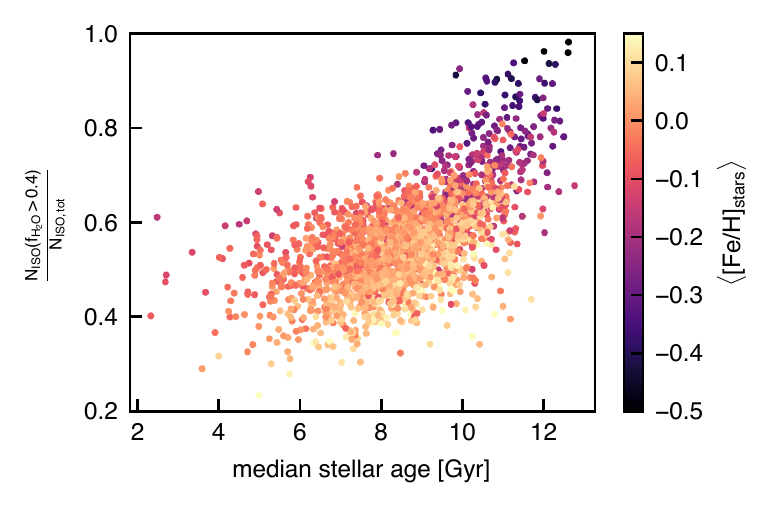}
    \caption{Water-rich ($f_{\mathrm{H_2O}} > 0.4$) ISO fraction as a function of galaxy median age for 2039 galaxies with $10^{10} < M_{\star} < 10^{11}\ \mathrm{M_{\odot}}$ from the Ref-L100N1504 EAGLE volume. Points are coloured by the mean stellar iron abundance relative to hydrogen, $\mathrm{[Fe/H]}$, as a proxy for the overall metallicity; as expected, this quantity is linked to the star formation history. There is a clear trend between median stellar age and water-rich ISO fraction, such that galaxies with older stellar populations harbour more water-rich ISOs. }
    \label{fig:population}
\end{figure}

In Figure \ref{fig:population}, we show the median stellar age against the water-rich ISO fraction for these galaxies, colouring the points by the mean stellar metallicity of each galaxy. There is a significant trend such that older, more metal-poor galaxies have more water-rich ISO populations. At intermediate ages, the trend is less significant, but the correlation with $\mathrm{[Fe/H]}$ demonstrates that strong constraints on the ISO population \emph{and} Galactic mean metallicity would provide a good constraint on the Galactic star formation history. The fact that even as simple a proxy for the star formation history as the median age of star formation correlates well with an observable property of the ISO population demonstrates how the observed population of ISOs in the Milky Way could constrain its early star formation history. In particular, we note that a low water-rich fraction is incompatible with a star formation history where the median age is greater than approximately 10Gyr. More complex modelling of -- and comparisons between -- the ISO population and detailed Galactic star formation history will be the objective of future work.

\subsection{Results for Milky Way analogue}

\begin{figure}
    \centering
    \includegraphics[width=\columnwidth]{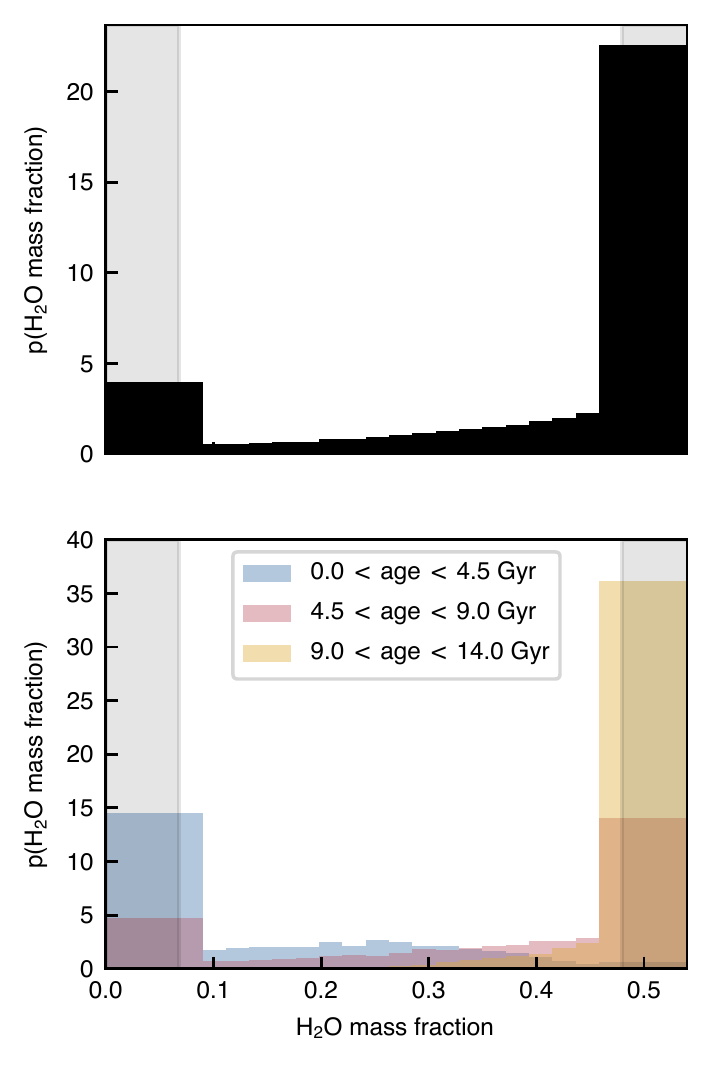}
    \caption{Top: Predicted distribution of ISO objects by water mass fraction. As before, the histograms are normalised such that $p(\mathrm{H_2O\ mass\ fraction})$ is the probability density for finding an ISO in a given mass fraction bin or equivalently the fraction of the predicted ISO population in each bin. The gray regions lie outside the formal limits of the BB20 model; here we assume that predicted ISOs have the minimum or maximum possible water mass fraction. Bottom: As above, but split by age of population. The low water mass fraction population is dominated by the youngest objects, while the population with high water mass fraction is from older systems.}
    \label{fig:my_label}
\end{figure}

We now consider predicting the ISO population for a single galaxy in the EAGLE simulation, the Milky Way analogue described in section \ref{sec:galmodel}. The results are shown in Figure \ref{fig:my_label}. ISOs display a bimodal distribution, with $60\%$ of the predicted population having $\mathrm{f_{H_2O}}>0.4$. A second substantial population is contributed by systems which form in higher metallicity conditions, which as noted earlier, have extremely low water fractions; $40\%$ have $\mathrm{f_{H_2O}}<0.4$. The populations with $\mathrm{f_{H_2O}} \gtrsim 0.4$ and $\mathrm{f_{H_2O}} \lesssim 0.1$ come from stellar systems with [Fe/H] $\lesssim$ -0.3 and [Fe/H] $\gtrsim$ 0.4, respectively. The model therefore predicts that, as more interstellar objects are observed, the population should be dominated by twin populations of high and low water mass fraction, with relatively few ISOs appearing with an intermediate water fraction. 

As expected from the enrichment history of the Milky Way, splitting the population by the age of the donor systems shows that this low water fraction population comes from younger stars (formed within the last 4~Gyr), whereas ISOs with higher water fractions and low CO mass fractions are dominated by those formed around older stars. The presence of a population of interstellar objects with low water fraction is therefore an indication of a contribution from recent star formation. In general, if the observed ratio of high to low water mass fraction interstellar objects varies from that predicted here, one must either adjust the adopted model of planetesimal formation, relax our assumption of rapid mixing, or change the details of the galactic evolutionary model. Recent work using data from Gaia (e.g. \citet{Mor:2019}) provide support for a model which suggests that the Milky Way has had a significant recent burst of star formation; our results suggest that observations of ISOs can provide a direct test of this idea.

\subsection{Comparison with observed ISOs}

Having derived a broadly bimodal population, comprising both high and low water fraction components, it is tempting to draw conclusions relative to the two interstellar objects observed so far. Solar System experience advises caution in translating the dust-to-gas mass ratios inferred from one-off comae observations into refractory-to-ice mass ratios in the body \citep{Choukroun:2020}. Considering \Ou\ as a planetesimal, 
with comet-coloured surface but effectively inactive \citep{Trilling:2018, ISSITeam:2019}, the tradeoffs in its dust-to-gas ratio vs its low bulk density leave its water fraction somewhat unconstrained \citep{Hui:2019}. 
2I/Borisov's coma showed water sublimation from at least 6 au, together with nitrogen and carbon \citep{Fitzsimmons:2018, Bannister:2020}.
It is  distinctively rich in CO relative to Solar System comets \citep{Bodewits:2020, Cordiner:2020}.
\citet{Seligman:2021} argue that both of these interstellar objects can be plausibly considered as CO-enriched planetesimals --- if 1I had variable activity, a common state for comets.

In the context of the BB20 models, stars of super-Solar metallicity have large carbon abundance, which binds into abundant CO and CO$_{2}$ exterior to the water line.
Such disks would prolifically produce CO-rich objects.
Consistently, the dynamical evidence suggests a young kinematic age for \Ou --- a mere 30-35 Myr, while 2I is dynamically hotter and thus older at around 700 Myr, though its age is not as well constrained \citep[e.g.][]{Hallatt:2020, Hsieh:2021}.
Thus, both observed ISOs would be from the putative recently formed, low water fraction population. 
One from the high water fraction population is yet to be seen. 
Speculatively, if ISOs with high water fractions form in low-metallicity disks, it may be that their dust:volatile ratio is different from their low water fraction counterparts; lacking dust, such objects may quickly sublime in the outer solar system, before they can be observed. Tempting though such speculation is, it is clear that drawing conclusions from two such briefly-observed objects is obviously premature. It is certain that more observations, over longer time-spans through perihelion, and best of all one day from spacecraft ground-truth (cf. Comet Interceptor \citep{Snodgrass:2019}), are needed to test these models.

Similarly, we suggest that information carried by the trajectories of observed interstellar objects might also be used to test the models. As an ISO travels through the Milky Way, it will be subject to on-average continuous dynamical heating, primarily from encounters with molecular clouds \citep{Pfalzner:2020} and dark matter substructure. Objects older than 1 Gyr will likely have been significantly heated vertically, and perhaps radially. An older population might thus be expected to have high velocity dispersion in all directions \citep{McGlynn:1989}; this line of argument, if sustained by further modelling, would predict that the more water-rich ISOs will have a more isotropic distribution of origins on the sky than objects with lower water fractions. Adding a detailed understanding of Milky Way dynamics, derived from \textit{Gaia} data, to our model will allow this prediction to be refined, in common with ISO trajectory expectations \citep{Seligman:2018}: this will lead to testable outcomes with the upcoming Vera C. Rubin Observatory sample. 

\citet{Gaidos:2017}, and most recently also \citet{Hallatt:2020} and \citet{Hsieh:2021} have suggested that the trajectory of \Ou\ was consistent with membership in the nearby Carina local group. 
If the first ISO observed does indeed have a local origin, this would suggests that our assumption here of thorough mixing through the Galactic disk may need revision. There are also additional complexities of density such as stellar streams and individual stars' ISO streams \citep[e.g.][]{Eubanks:2019, Portegies:2021streams}.

However, this assignment of origin seems to also require unfeasible numbers of ISOs to be donated by each Carina stellar system, relative to the local solar ISO spatial density constraint. \citet{Pfalzner:2021clouds} show that molecular clouds are regions of dense capture-and-release for interstellar objects. 
It is plausible that not all ISOs tracing to Carina originate in Carina stellar systems, and the molecular cloud is locally overdense with ISOs, perhaps as a consequence of cloud formation (c.f. \citet{Pfalzner:2019,MoroMartin:2021}).

\section{Conclusions}

This paper brings together two very different modelling efforts. The first, by \citet{Bitsch:2020}, predicts the composition of planetesimals formed beyond the ice line in a protoplanetary disk, and the second predicts the distribution of metallicity in Milky Way analogues from the EAGLE simulations \citep{Schaye:2015,Crain:2015}. We combine these efforts to predict the water mass fraction distribution for a population of interstellar objects, assuming that the resulting population is well mixed and sampled randomly by the motion of our Solar System through the galaxy. 

The bulk of the resulting predicted interstellar object population has high ([Fe/H]$>$0.4) water mass fractions, primarily from older systems which formed more than 4.5 Gyr ago. A second, smaller but still substantial, population has low ([Fe/H]$<$0.1) water mass fractions and is primarily due to more recent star formation; both interstellar objects seen thus far appear to belong to this population. As more ISOs are discovered, the relative frequency of those with high and low water fractions will provide a constraint on the effectiveness of mixing between old and young populations in the Milky Way, and this parameter also shows promise as a constraint on the Milky Way's star formation history. Such constraints from models could be compared with studies of stellar populations, providing tests of the same models with a different selection function. The ISO population will contain small bodies whose origins lie in systems where the central star has since died. Thus, studying such objects affords a different sampling of galactic star formation history than that possible from studies of extant stars.

This necessarily preliminary study demonstrates the ability to make specific predictions about the nature of the interstellar object population from existing models. Extensions which consider the abundances of other molecules will add to the predictive power of our modelling. Further work which uses large scale simulations such as EAGLE to provide insight into the nature of the ISO population is encouraged, and will be needed ahead of the beginning of the LSST to understand how properties of the observed population can be used to test the preliminary predictions presented here.

\acknowledgments

We thank the reviewer for their comments, which greatly helped in clarifying the arguments presented in this paper.

This paper benefited from much-appreciated discussions with Matthew Hopkins, and the International Space Science Institute (ISSI) ‘Oumuamua team in Bern, Switzerland, Susanne Pfalzner, Dennis Bodewits, Max Briel and Jan Eldridge, Joe Masiero, and Bruce Macintosh.

MTB appreciates support by the Rutherford Discovery Fellowships from New Zealand Government funding, administered by the Royal Society Te Ap\={a}rangi.

\software{Astropy \citep{Astropy:2018}, NumPy \citep{Numpy:2020}  
          }

\bibliography{references}{}
\bibliographystyle{aasjournal}


\end{document}